\documentclass[preprint2]{aastex}

\newcommand{\sst}{{\it Spitzer}} 

\slugcomment{}

\shorttitle{Spitzer Space Telescope FLS R-band Images.}
\shortauthors{Fadda et al.}

\begin{document}


\title{The Spitzer Space Telescope First-Look Survey: \\ 
KPNO MOSAIC-1 R-band Images and Source Catalogs}


\author{Dario Fadda\altaffilmark{1}, Buell T. Jannuzi\altaffilmark{2},
Alyson Ford\altaffilmark{2}, Lisa J. Storrie-Lombardi\altaffilmark{1}}


\altaffiltext{1}{Spitzer Science Center, California Institute of Technology, MS 220-6, Pasadena, CA 91125}
\altaffiltext{2}{National Optical Astronomy Observatory, Tucson, AZ 85719}


\begin{abstract}


We present R-band images covering more than 11 square degrees of 
sky that were obtained in preparation for the 
Spitzer Space Telescope First Look Survey (FLS). The FLS was designed
to characterize the mid-infrared sky at depths 2 orders of magnitude
deeper than previous surveys.  The extragalactic component is the first
cosmological survey done with \sst.  Source catalogs
extracted from the R-band images are also presented.  The R-band
images were obtained using the MOSAIC-1 camera on the Mayall 4m
telescope of the Kitt Peak National Observatory.  Two relatively large
regions of the sky were observed to modest depth: the main FLS extra
galactic field ($17^h18^m00^s$ $+59^o30'00''.0$ J2000; l$=88.3$,
b$=+34.9$) and ELAIS-N1 field ($16^h10^m01^s$ $+54^o30'36''.0$ J2000;
l$=84.2$, b$=+44.9$).  While both of these fields were in early plans
for the FLS, only a single deep pointing test observation was made
at the ELAIS-N1 location. The larger Legacy program SWIRE (Lonsdale et
al., 2003) will include this region among its surveyed areas.  The
data products of our KPNO imaging (images and object catalogs) are
made available to the community through the World Wide Web (via the
Spitzer Science Center and NOAO Science Archives).  The overall quality 
of the images is high. The measured positions of sources detected in the images have
RMS uncertainties in their absolute positions of order 0.35
arc-seconds with possible systematic offsets of order 0.1 arc-seconds,
depending on the reference frame of comparison.  The relative
astrometric accuracy is much better than 0.1 of an arc-second.
Typical delivered image quality in the images is 1.1 arc-seconds full
width at half maximum.  Images are relatively deep since they reach a
median 5$\sigma$ depth limiting magnitude of R=25.5 (Vega), as
measured within a 1.35~$\times$~FWHM aperture for which the S/N ratio
is maximal.  Catalogs have been extracted using SExtractor using
thresholds in area and flux for which the number of false detections
is below 1\% at R=25. Only sources with S/N greater than 3 have been
retained in the final catalogs.  Comparing the galaxy number counts
from our images with those of deeper R-band surveys, we estimate that
our observations are 50\% complete at R=24.5. These limits in depth
are sufficient to identify a substantial fraction of the infrared
sources which will be detected by {\it Spitzer}.

\end{abstract}


\keywords{astronomical data bases:catalogs --- galaxies: photometry}

\section{Introduction}

One of the main advantages of the Spitzer Space Telescope (formerly
known as the Space Infrared Infrared Telescope Facility; SIRTF,
Gallagher et al. 2003) is the possibility to make extragalactic
surveys of large regions of the sky in a relatively short time
covering wavelengths from the near-IR to the far-IR with the
instruments IRAC (Fazio et al. 1998) and MIPS (Rieke et al. 1996).
Compared to {\it Spitzer}'s predecessors (e.g. IRAS, Soifer et~al. 1983, and ISO,
Kessler et al. 1996), there are improvements in the detectors
(number of pixels and better responsivity), the collecting area of the
primary mirror (85cm diameter), and Sun-Earth-Moon avoidance
constraints due {\it Spitzer}'s heliocentric orbit.  
\sst\ can also make observations simultaneously in multiple 
bands (with IRAC--3.6 and 4.5 or 5.8 and 8 $\mu$m; with MIPS--24, 70 and 160 $\mu$m).

Many extragalactic surveys are already scheduled with {\it Spitzer} as
Legacy programs (SWIRE, Lonsdale et al. 2003; GOODS, Dickinson \&
Giavalisco 2001) or as observations by the Instrument Teams
(Wide, Deep and Ultra-Deep {\it Spitzer} surveys which will cover regions like
the Bo\"otes field of the NOAO Deep Wide-Field Survey, the Groth
strip, Lockman Hole, XMM-Deep and so on, see e.g.~Dole et al. 2001).
The First Look Survey utilizes 112 hours of Director's Discretionary 
time on \sst\ and includes extragalactic, galactic, and ecliptic
components\footnote{see the FLS website at ssc.spitzer.caltech.edu/fls}. These 
data will be available to all observers when the Spitzer Science 
Archive opens in May, 2004. The purpose of the FLS is to characterize
the mid-infrared sky at previously unexplored depths and make these
data rapidly available to the astronomical community. 
The extragalactic component is comprised of a 4 square degree
survey with IRAC and MIPS near the north ecliptic pole centered at
J1718+5930. These observations were executed on 2003 December 1-11.

To fully exploit the \sst\ FLS data we have obtained ancillary surveys 
at optical (this paper) and radio wavelengths (Condon et~al. 2003).  
Given the modest spatial resolution of the {\it Spitzer} imagers (the point
spread function is large, especially in the mid and far-IR, e.g. 5.7
arcseconds full width at half maximum, FWHM, for the 24 $\mu$m
channel), the first problem to solve for the infrared sources detected
by {\it Spitzer} will be to associate these sources with an optical
counterpart, when possible. This will then allow the higher spatial
resolution of the available optical images to assist with the source
classification (e.g. as stars, galaxies or QSO) and enable
targeting of subsets of the sources for spectroscopy with optical or
near-IR spectrographs.
Since many of the infrared sources that will be detected by {\it Spitzer} will
be dust-obscured galaxies with faint optical counterparts, the
complementary optical imaging must be relatively deep.

Although a deep multi-wavelength optical survey would be more useful,
allowing one to compute photometric redshifts (e.g. the NOAO Deep
Wide-Field Survey, Jannuzi and Dey 1999, Brown et al. 2003), the task
of deeply covering a large region of sky in a homogeneous manner is
quite time-consuming. Therefore, for the initial optical ancillary survey 
we chose to observe the entire field in the R-band.  NOAO provided 
4 nights of Director's Discretionary time on the KPNO 4m in May, 2000, for
this survey. We have limited multi-wavelength optical observations to 
the central portion of the
FLS field. The Sloan Digital Sky Survey included the FLS field in their 
early release observations (Stoughton et al. 2002) and mosaics and catalogs
for the region are also now available (Hogg et~al. 2004). 

In this paper, we present the R-band optical observations made with
the MOSAIC-1 camera on the Mayall 4m Telescope of Kitt Peak National
Observatory. Centered on the main FLS field, a region 9.4 square
degrees in area was imaged. In addition, 2.3 square degrees covering
the ELAIS-N1 field was also observed.  Although originally the
ELAIS-N1 field was planned to be part of the FLS program, the FLS
observations of the ELAIS-N1 field have now been revised to a very
deep 10'$\times$10' pointing to evaluate the confusion limits of the
MIPS instrument.  The remainder of the ELAIS-N1 field will now be
imaged as part of a larger survey in this region, a portion of the SWIRE
{\it Spitzer} Legacy Survey (Lonsdale et al., 2003).

In section 2 we review the overall observing strategy and describe the
MOSAIC-1 observations. In section 3 we discuss the techniques used in
the data reduction including the astrometric and photometric
calibration of the images. We describe in section 4 the data products
made publicly available.  We detail in section 5 the criteria used to
detect, classify, and photometrically measure objects in the images.
Section 5 also includes a description of the information available in
our catalogs.  In section 6 we examine the quality of the imaging data
by comparing them with other available data sets. Finally, a brief
summary is given in section 7.

\section{Observations}

\begin{figure}[!t]
\centerline{\includegraphics[width=0.5\textwidth,angle=0]{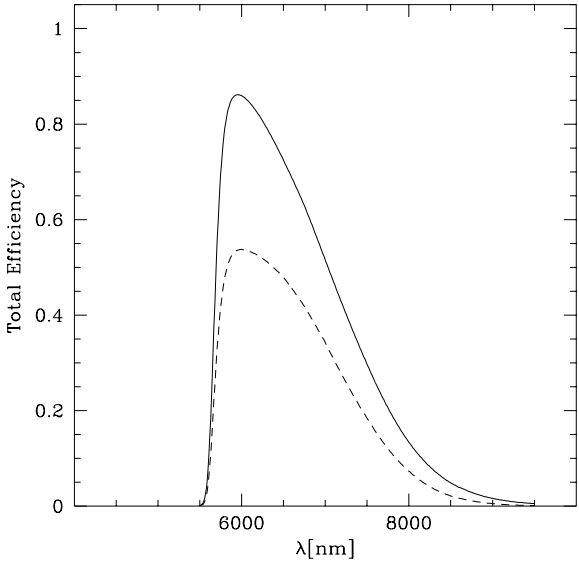}}
\caption{Transmission curve of the R-band filter (solid line) used for
the observations.  The dashed line indicates the combined response
when considering also the CCD quantum efficiency, the throughput of
the prime focus corrector, 
and the atmospheric absorption with a typical airmass of 1.2.  }
\label{fig:transcurve}
\end{figure}

The optical observations of the FLS region (centered at $17^h18^m00^s$
$+59^o30'00''$, J2000) and of the ELAIS-N1 region (centered at $16^h10^m01^s$
$+54^o30'36''$, J2000, see Oliver et al. 2000) were carried out using the
MOSAIC-1 camera on the Mayall 4m Telescope at the Kitt Peak National
Observatory. The MOSAIC-1 camera is comprised of eight thinned,
back-illuminated SITe 2048 $\times$ 4098 CCDs with a projected pixel
size of 0.258 arcsec (Muller et al.~1998).  The 8 CCDs are physically
separated by gaps with widths of approximately 14 and 15.5 arcsec
along the right ascension and declination directions, respectively.
The full field of view of the camera is therefore 36 $\times$ 36
square arcmin, with a filling factor of 97\%.  Observations were
performed using the Harris Set Kron-Cousins R-band filter whose main
features are summarized in Table~\ref{tbl:filter}. The transmission
curve of the filter is shown in Figure~\ref{fig:transcurve} together
with the resulting modifications that would be introduced by the
corrector and camera optics, the CCD quantum efficiency, and
atmospheric extinction.

For organizational purposes we chose to divide the proposed FLS survey
region into 30 subfields each roughly the size of an individual
MOSAIC-1 pointing.  The coordinates of these subfields are listed in
Table~\ref{tbl:coordinates}.  During our observing run we were able to
complete observations for 26 of these subfields. We similarly divided
the ELAIS-N1 field into 12 subfields, but only completed observations
for 5 of these fields.  Each subfield was observed for a minimum of
three 10 min exposures.  In practice, some images were not suitable
(poor seeing, flat fielding problem, or some other defect) and were
not included in the final coadded or ``stacked'' images we are
providing to the community. Table~\ref{tbl:coordinates} lists in
Col. 4 the number of exposures which were obtained and included in the
coadded or ``stacked'' images.  In order to provide some coverage in
the regions of the sky that would fall in the inter-chip gaps, the
positions of successive exposures of a given field were offset on
order of an arcminute relative to each other.  In general the first
exposure was at the nominal (tabulated) position, the second
shifted by $41.5'$ in $\alpha$ and $-62.3'$ in $\delta$, and the third with a
shift of $-41.5'$ and 62.3$'$ in $\alpha$ and $\delta$.  In a
few cases an additional position with $\Delta\alpha = 20.8'$ and
$\Delta\beta=31.1'$ has been observed. For a few fields, observations
have been repeated due to the bad seeing or pointing errors.

The KPNO imaging observations were made on 2000 May 4-7,9 UT.  A log
of the observations is summarized in Table~\ref{tbl:log}, which lists
for each group of observations: the date of the observations (Col. 1);
the subfield name (Col.2); the integration
time and the number of exposures (Col. 3); and the seeing (delivered
image quality expressed as FWHM in arcseconds of bright unsaturated
stars) range of each exposure (Col. 4).

\begin{figure}[t]
\centerline{\includegraphics[width=0.5\textwidth,angle=0]{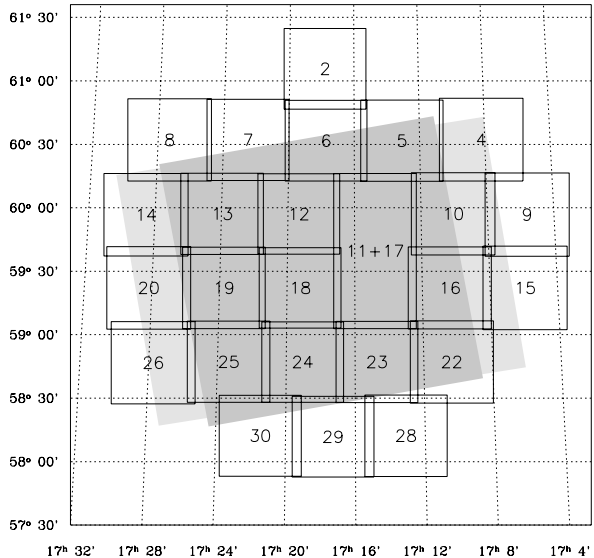}}
\caption{The KPNO fields in the FLS region cover
most of the MIPS (grey) and IRAC (dark grey) Spitzer
observations.}
\label{fig:FLSfields}
\end{figure}

\begin{figure}[!b]
\centerline{\includegraphics[width=0.5\textwidth,angle=0]{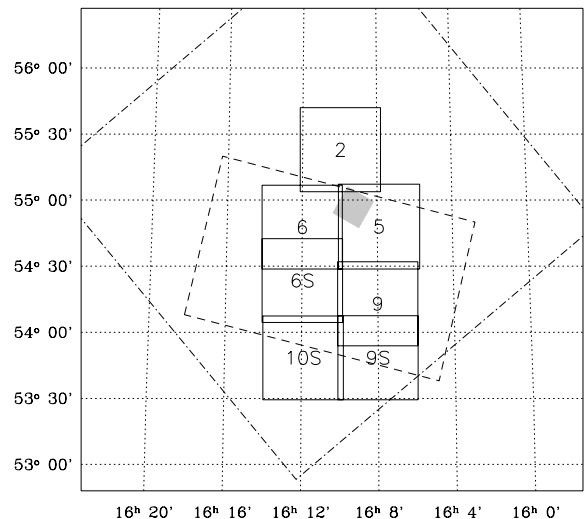}}
\caption{The KPNO fields in the ELAIS-N1 region. The dashed line
corresponds to the region observed by ISOCAM at 14.3 $\mu$m (Oliver et
al. 2000). The grey shaded square is the field observed with Spitzer
to test the MIPS 24 micron confusion limit. The SWIRE planned field
(Lonsdale et al. 2003) is marked with a dash-dotted line. }
\label{fig:ELAISfields}
\end{figure}

During a portion of the observing run, the pointing of the telescope
was incorrectly initialized, resulting in approximately a $24'$ error
in the pointing for some fields. These are noted in the observing log.
Since the $24'$ offset did not map exactly on to our subfield grid,
we chose to make stacked or combined images for several of the
subfields.  This was done for subfields \#~11 and \#~17 in the FLS
region and subfields (\#~6, \#~9 and \#~10) of ELAIS-N1.
In Figures~\ref{fig:FLSfields} and~\ref{fig:ELAISfields} we show the
positions of each subfield on the sky with respect to the fields
which have been covered with the IRAC and MIPS instruments on-board
{\it Spitzer}. In Figure~\ref{fig:ELAISfields} we display the position of the
subfields in the ELAIS-N1 region and the region observed with ISOCAM
as part of the European Large Area ISO Survey (ELAIS, Oliver et
al. 2000). The shaded square indicates the area which has been observed 
to test the confusion MIPS 24 micron confusion limits, as part of the FLS
observations.

\section{Data Reduction}
\subsection{Basic Reductions}

The processing of the raw MOSAIC-1 exposures followed the steps
outlined in version 7.01 of ``The NOAO Deep Wide-Field Survey MOSAIC
Data Reductions
Guide''\footnote{www.noao.edu/noao/noaodeep/ReductionOpt/frames.html}
and discussed by Jannuzi et al. (in preparation).  The bulk of the
software used to process the images and generate combined images for
each subfield from the individual 10 minute exposures is described by
Valdes (2002) and contained as part of the MSCRED software package
(v4.7), which is part of IRAF.\footnote{IRAF is distributed by the
National Optical Astronomy Observatory, which is operated by the
Association of Universities for Research in Astronomy, Inc., under a
cooperative agreement with the National Science Foundation}

The image quality of the final stacks is variable, as was the seeing
during the run. Users of the images should be aware that the detailed
shape of the point spread function in a given image stack could be
variable across the field, not only because of residual distortions in
the camera, but because a given position in the field might be the
average of different input images, each with their own point spread
function. No attempt was made to match the PSFs of the individual
images before combining the images.

In a survey area this large there will be fields with very
bright stars. This can cause some regions of the survey area
to be impacted by scattered light from these stars. Some effort
was made to minimize this impact during the observations (by
shifting pointings) and reduction of the images (through
masking of affected regions in some images to allow unaffected
images to be the sole contribution to the stacked image), but
users of the images should be aware that some scattered
light will have made it into some of the stacks. An example of
a subfield with significant scattered light is FLS\_15.

Flat fielding of the images was accomplished through the application
of calibration files generated first from observations of a flat-field
screen inside the dome at the 4m and from a ``super-sky'' flat
constructed by combining the majority of the FLS and ELAIS-N1 images
(with objects masked and rejected).  The result is generally excellent
flat fielding of the sky (with some fields with slightly different sky
color, due to moon light or twilight worse), but users should be aware
that there are likely to be some color-dependent variations between
and within the 8 CCDs, meaning that uncorrected errors in the
photometry of a given object, attributable to the flat fielding
correction not being derived from a source with color matched to the
color of that object, on the order of 1 to 3 percent could still be
present in the data even though the sky is generally quite ``flat''.

\subsection{Astrometric Calibration}

The astrometric calibration of the MOSAIC-1 images is accomplished in
two steps. First, the high-order distortions in the field, which will
in general be common to all our exposures (those anticipated to be the
result of the optics and/or CCD placement and as a result stable over
the course of an observing run or season) are calibrated using
observations of an astrometric standard field.  
These distortions can be wavelength dependent, so the calibration is made for the specific
filter being used.  These images are analyzed to produce a default
correction for each of the CCDs.  Low-order corrections (translational
offset, small rotation, and/or scale adjustments needed to compensate
for pointing errors, instrument mounting variations between runs, and
atmospheric effects) are corrected on an exposure by exposure basis
using the many catalog objects in each exposure, the previously
mentioned knowledge of the high-order distortions as a function of CCD
position, and the software {\it msccmatch} in IRAF.  
The astrometric calibration has been performed using the reduction protocol
developed for the reduction of the NOAO Deep Wide-Field Survey (NDWFS) data
which assumes as astrometric reference the  USNO-A2.0 catalog (Monet et~al.~1998).
Adopting the GSC~II catalog,\footnote{The Guide Star
Catalogue-II is a joint project of the Space Telescope Science
Institute and the Osservatorio Astronomico di Torino. Space Telescope
Science Institute is operated by the Association of Universities for
Research in Astronomy, for the National Aeronautics and Space
Administration under contract NAS5-26555.  The participation of the
Osservatorio Astronomico di Torino is supported by the Italian Council
for Research in Astronomy.  Additional support is provided by European
Southern Observatory, Space Telescope European Coordinating Facility,
the International GEMINI project and the European Space Agency
Astrophysics Division.}  which became available after the development
of the NDWFS reduction protocol, might improve the quality of the
astrometric solutions and this might be done if a rereduction of the
data set is done in the future.  As we discuss further below, the
anticipated improvement would be slight.  For the images described in
this paper our solutions for the mapping of the MOSAIC-1 pixels into a
world coordinate frame had an RMS scatter of between 0.3 to 0.45
arcseconds, depending on the particular field.

After each image was provided with an improved astrometric
calibration, it was tangent-plane projected with respect
to the FLS and ELAIS-N1 field center positions listed above.
Following projection, the individual exposures (typically three for
a given region of the sky) could be combined into a final, stacked,
image.  These are the images currently being made available in
the {\it Spitzer} and NOAO archives.

During the final stages of the data reduction, when the individual
images were being combined to make the final image stacks, it was
determined that there was an error in the high-order astrometric
correction file for the Harris R-band filter affecting the adjacent
edges of CCDs 7 and 8 in the MOSAIC-1 camera.  The original solutions
for the high order distortion terms were determined on a chip by chip
basis with no requirement that the solution be continuous across the
entire field. In general, while this requirement was not imposed, it
was met by the solutions provided by NOAO.  However, for the R-band
the solution available at the time we were reducing the data was
discontinuous at the CCD7 and CCD8 boundary.  Since our image stacks
are made from the combination of three or more images that are offset
by 30 arcseconds to an arcminute from each other, this difference
between the solutions for the two CCDs can result in a mis-mapping,
into RA and DEC space, of a region of the sky imaged first on CCD7 and
then on CCD8.  The error is small, and introduces at most an
additional $0.1''$ uncertainty to the positions of sources in the
affected region (which is located in the SE Corner of each field,
about 25\% of the field up from the southern edge), but will result in
a degradation of the PSF for objects affected in this region. The size
of the affected region in each subfield is approximately 8.5$'$
East-West and 2.5$'$ North-South, or a bit less than 2\% of the
surveyed area.

\subsection{Photometric calibration}

\begin{figure} [b!]
\centerline{\includegraphics[width=0.5\textwidth,angle=0]{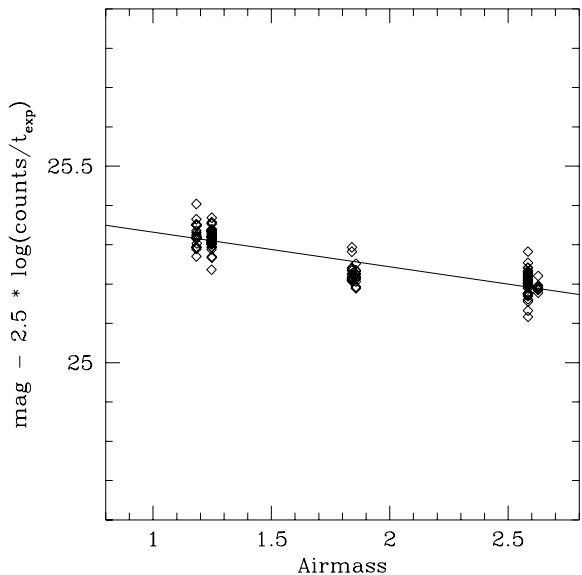}}
\caption{Dependence of zero-point on the airmass. Each squares is
a standard star used in the calibration with 9 standard fields observed
during the night of 2000 May 5.} 
\label{fig:zeropoint}
\end{figure}

Since not all the subfields were observed during nights with
photometric conditions (see Table~\ref{tbl:log}), we derived a coherent photometric system in
two steps. First, we computed the relative photometric zero-points
between the different stacked images of the subfields.  Since each
subfield overlaps its neighbor by approximately two arcminutes, we
can estimate the extinction difference between them using a set of
common sources.  This can be expressed in terms of zero-point that, by
definition, includes the effects of airmass and extinction.  Because
each frame has multiple overlaps, the number of frame-to-frame
magnitude differences is over-determined with respect to the number of
frames.  Therefore, one can derive the relative zero-point for each
frame simultaneously using a least squares estimator. 
We extracted the sources from each subfield using as a first guess for
the zero-point that we computed for a central subfield (\#18
and \#5 for the FLS and ELAIS-N1 fields, respectively).  For each
overlap between contiguous fields, we selected the pairs of stars with
magnitude $18<R<21$ and computed the median of the difference in
magnitudes by using a 3$\sigma$ clipping procedure.  The magnitudes
considered were the auto magnitudes (MAG\_AUTO) from SExtractor, which are fairly
robust with respect to seeing variations.  To estimate the relative
zero-points in the sense of the least squares, we minimized the sum:
\begin{equation}
\sum_{i > j} N_{ij}^2 (z_{i} - z_{j} - \Delta_{ij})^2,
\end{equation}
where $z_i$ is the variation  with respect to the
initial guess of the zero-point of the subfield $i$, and $\Delta_{ij}$ is the median of
the differences of magnitude for the set of the $N_{ij}$ source pairs in
the overlapping region between the subfields $i$ and $j$.
Solving the linear system obtained by imposing that the derivatives of
(1) with respect each $z_i$ are equal to zero, we corrected the initial
guesses for the zero-points. The procedure was then iterated until the
number of pairs $N_{ij}$ became stable.

The second step was to make use of those fields observed under
photometric conditions in order to converge on the best
zero-point. Standards were measured several times over a broad
range of airmasses on the first two nights of the observations, which
had the best photometric conditions (see Table~\ref{tbl:log}). Only the second night (May 5) was
really photometric since all the measurements are coherent (see
Figure~\ref{fig:zeropoint}).  Magnitudes were calibrated to the
Kron-Cousins system using Landolt standards taken from Landolt
(1992). They have been obtained using an aperture of 6 arcsec in
diameter, large enough to obtain accurate measurements according to
the growth curves of all the measured stars. We have fitted the linear
relationship:
\begin{equation}
m = m_0 - 2.5 \log (C/t_{exp}) + m_X * A,
\end{equation}
with $m$ magnitude of Landolt standard, $A$ airmass during the
observation, $C$ counts during the exposure time $t_{exp}$, $m_0$  
and $m_X$ the zero and extinction terms, respectively.
We used an iterative 3$\sigma$ clipping in order to discard deviant
measurements, finding the best fit for  $m_0 = 25.42$ and $m_X = -0.09$.
The standard deviation of the residuals for the best fit is 0.03.

\section{Data Products}

The analysis of the survey data produced a set of intermediate and
final products, images and catalogs, which are publicly available at
the Spitzer Science Center and NOAO Science Archives\footnote{ssc.spitzer.caltech.edu/fls/extragal/noaor.html and www.noao.edu}.  
In particular, we provide the astronomical community with:

{\bf Coadded (Stacked) Images of the Subfields}: sky-subtracted, fully
processed coadded frames for each subfield. The
subfields are mapped using a tangential projection. The size of each
fits compressed image is 220 MB. For each image, a bad pixel mask and
an exposure map is given (in the pixel-list IRAF format).  The
photometric zero-point of each subfield after the absolute photometric
calibration of the frame appears in the header of the image.  A full
description of the header keywords is available at the http sites.

{\bf Low-resolution field image}: sky-subtracted, fully processed
coadded images of the whole field. The images have been created to
give a general overview of the FLS and ELAIS-N1 fields and have been
produced using a 1.3 arcsec pixel size (corresponding to five times
the original size of the pixel). Users are discouraged from using them
to extract sources and compute photometry.  These images have a size of
approximatively 380 Mb and 80 Mb.

{\bf Single subfield catalogs}: object catalogs associated with each single
subfield. A full description of the parameters available is described in the
following Section. The catalogs are in ASCII format.

\section{Catalogs}

The main goal of this survey is to catalog galaxies and faint stars
and make a first distinction between stars and galaxies on the basis
of their intensity profiles. Several bright objects (mainly stars) are
saturated and excluded from the catalog, but can be found in catalogs
from shallower surveys (like the Sloan Digital Sky Survey in the case
of the FLS region, see e.g. Stoughton et al. 2002 and Hogg et~al. 2004). 
The source extraction was performed with the SExtractor package (Bertin
\& Arnouts 1996; ver. 2.3) which is well suited for surveys with
low to moderate source density as is the case of our surveys.

\subsection{Detection}

Several parameters have to be fixed to achieve an efficient source
extraction with SExtractor.  The first problem is the evaluation of
the background. SExtractor proceeds computing a {\sl mini-background}
on a scale large enough to contain several faint objects and filtering
it with a box-car to avoid the contamination by isolated, extended
objects.  Finally, a full-resolution background map is obtained by
interpolation and it is subtracted from the science image.  In our
case, many bright stars populate the FLS field since it has a moderate
galactic latitude (34.9 degrees) while the problem is less important
in the case of the ELAIS-N1 field (gal. latitude: 44.9 degrees).
To evaluate a background which is not locally dominated by bright
stars we adopted meshes of 128$\times$128 pixels for the {\sl mini-background}
corresponding to 33.0 arcsec and used a 9$\times$9 box-car for the
median-filtering.

In order to improve the detection of faint sources, the image is
filtered to enhance the spatial frequency typical of the sources with
respect to those of the background noise. A Gaussian filter with an
FWHM similar to the seeing of the image (in our case 4 pixels, since
the overall seeing is 1.1) has been used. Although the choice of a
convolution kernel with a constant FWHM may not always be optimal
since the seeing is varying in the different images, the impact on
detectability is fairly small (Irwin 1985). Moreover, it has the
advantage of requiring no changes of the relative detection threshold.

Finally, the detection is made on the background-subtracted and
filtered image looking for groups of connected pixel above the
detection threshold.  Thresholding is in fact the most efficient way
to detect low surface brightness objects.
In our case, we fixed the minimum number of connected pixels to 15 and
the detection threshold to 0.8 (in units of the standard deviation of the
background noise), which corresponds to a typical limiting
surface brightness $\mu_{R} \sim 26$ mag arcsec$^{-2}$. 
For the detection we made use also of the exposure map as a weight considered
to set the noise level for each pixel. Some pixels have a null weight since
they correspond to saturated objects, trails of bright objects and other
artifacts. These pixels are also marked in the bad pixel mask and the
false detections around these image artifacts are flagged and easily 
excluded from our final catalogs.

Only objects detected with a signal-to-noise greater than 3 (based on the
total magnitude errors) are accepted in our final catalog.

Although not well suited to detect objects in crowded fields,
SExtractor allows one also to deblend close objects using a multiple
isophotal analysis technique. Two parameters affect the deblending:
the number of thresholds used to split a set of connected pixels
according to their luminosity peaks and the minimal contrast (light in
a peak divided by the total light in the object) used to decide if
deblending a sub-object from the rest of the object. In our analysis
we used a high number of thresholds (64) and a very
low minimal contrast (1.5 $e^{-5}$). Nevertheless, a few blended
objects still remain in the catalogs. Visual inspection or other
extraction algorithms more efficient in crowded fields (e.g, DAOPHOT)
are needed to treat these particular cases.

\subsection{Photometry}

The photometry has been performed on the stacked images. Several
measurements were made: aperture and isophotal magnitudes and an
estimate of the total magnitudes. We measured the aperture magnitude
within a diameter of 3 arcsec, roughly corresponding to three times
the overall seeing.  Total magnitudes ({\sl MAG\_AUTO}) are estimated
using an elliptical aperture with an approach similar to that proposed
by Kron (1980).
Since these fields have been selected in sky regions with low Galactic
extinction to observe extragalactic infrared sources, the corrections
for Galactic extinction (Schlegel et al. 1998) are small:  0.06 and 0.01
on average for the FLS and ELAIS-N1 fields, respectively.

\begin{figure*}
\centerline{\includegraphics[width=\textwidth,angle=0]{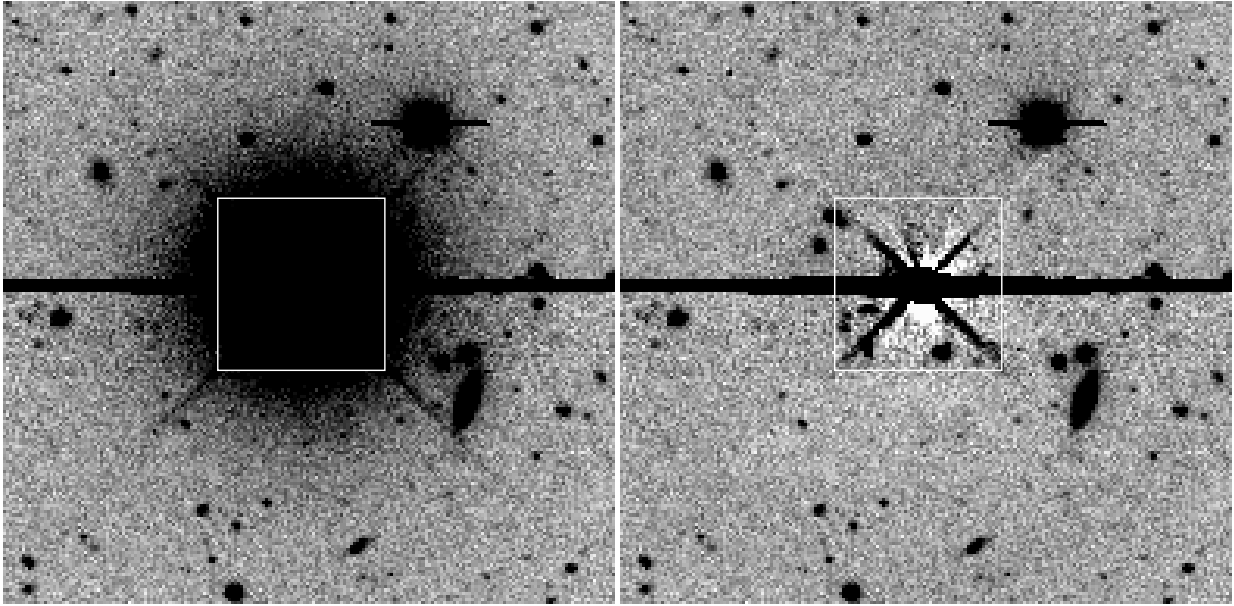}}
\caption{A 1.5'$\times$1.5' field around a bright saturated star in the field FLS\_6
before and after the star removal. The overplotted square delimits the
region which is not considered in the final catalog.}
\label{fig:removestar6}
\end{figure*}
\begin{figure*}
\centerline{\includegraphics[width=\textwidth,angle=0]{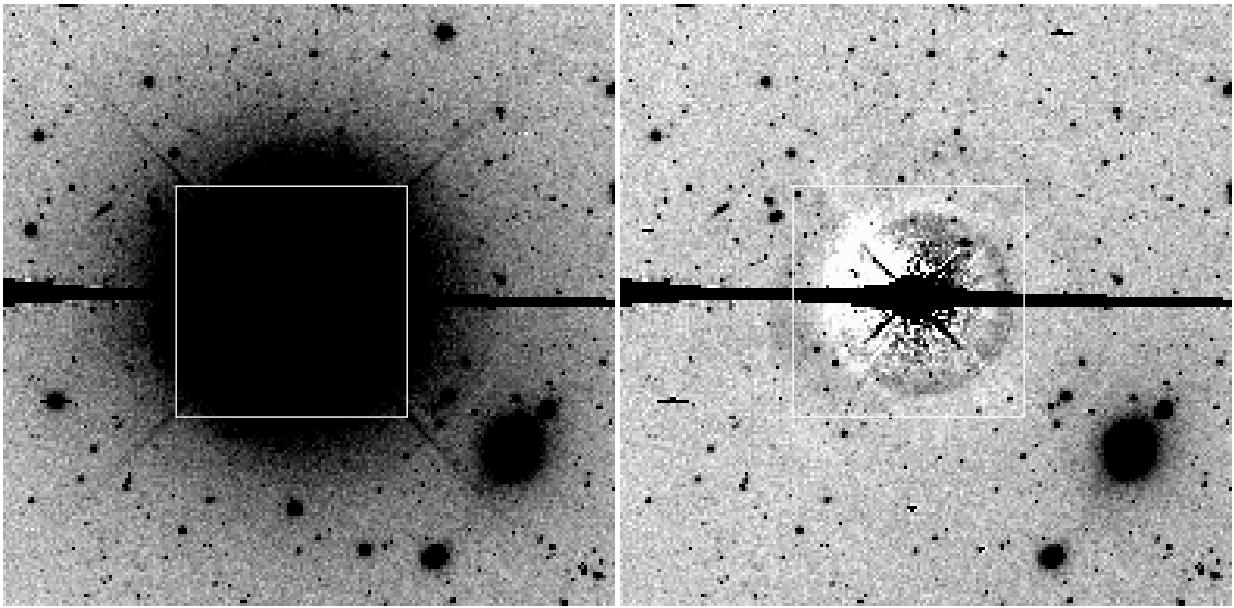}}
\caption{A 7'$\times$7' field around a bright saturated star in the field FLS\_13
before and after the star removal. The overplotted square delimits the
region which is not considered in the final catalog. Multiple reflections of the star
between the optics and the CCD are visible.}
\label{fig:removestar13}
\end{figure*}

The total magnitude of sources close to bright objects are usually
inaccurate since the local background is affected by the halo of the
bright objects and the Kron radius is not correctly computed.  To
improve the photometry for these sources we have subtracted bright
saturated stars from the images and excluded from the catalogs the sources
detected in square boxes around these stars where the subtraction is
not correct. Moreover, we have considered bright extended galaxies
and excluded from the catalogs all the sources inside the Kron ellipses
of the galaxies. In fact, most of these sources are bright regions of the
galaxies or their photometry is highly affected by the diffuse luminosity
of the galaxies.

To subtract bright saturated stars from the images we have computed
radial density profiles on concentric annuli around the stars. Then,
after subtracting these profiles, we have removed the diffraction
spikes by fitting their profiles along the radius at different angles
with Chebyshev polynomials. As visible in Figures~\ref{fig:removestar6}
and~\ref{fig:removestar13}, the background is much more uniform and
the spikes become shorter. This improves the photometry for the objects
surrounding the stars and avoid the detection of faint false sources
on the diffraction spikes.

\begin{figure}[ht!]
\centerline{\includegraphics[width=0.5\textwidth,angle=0]{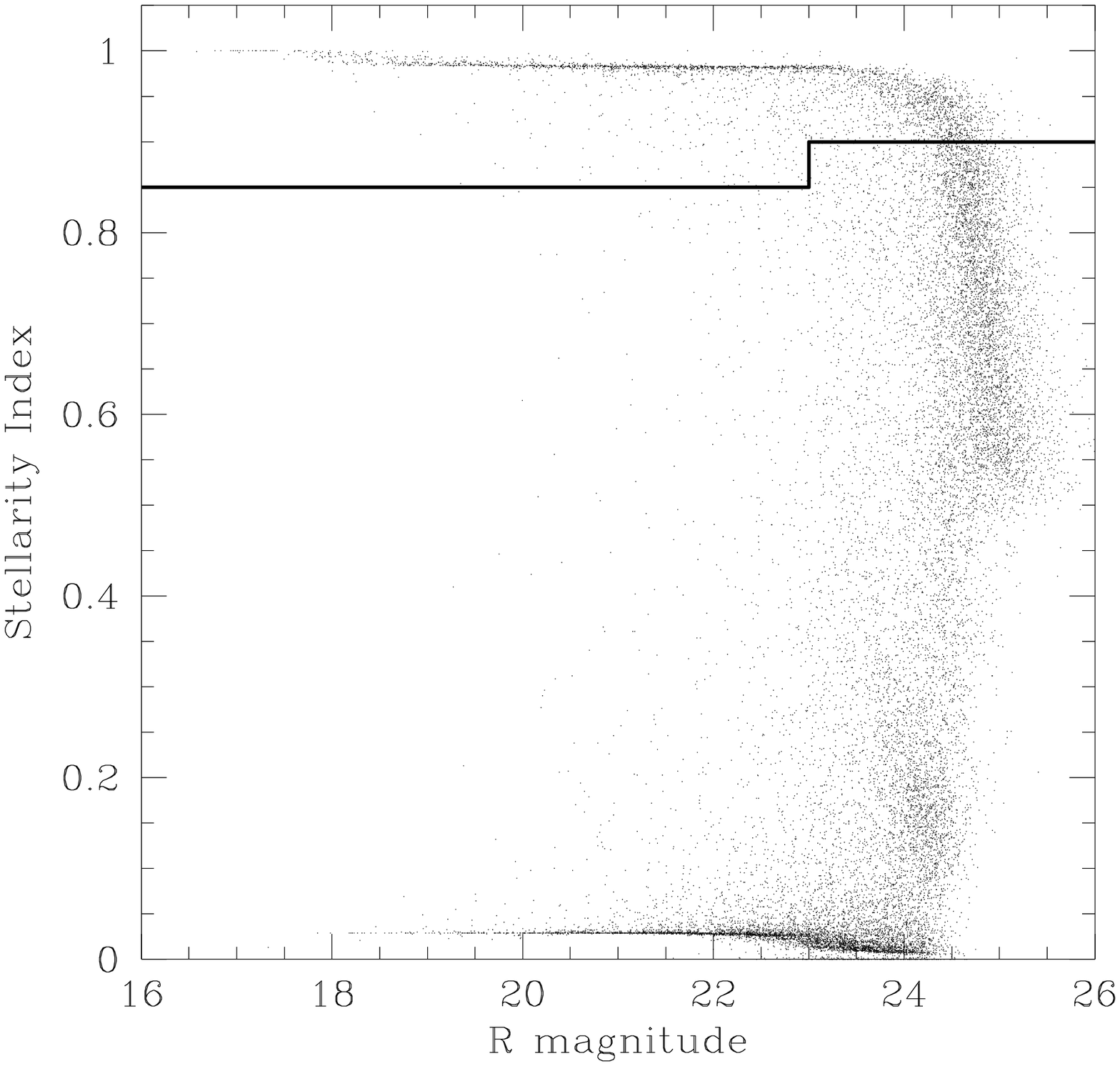}}
\caption{SExtractor stellarity index versus R magnitude for the objects
detected in the field FLS18. The solid line indicates the threshold of the 
stellarity index as a function of the magnitude chosen to separate stars from
galaxies in our survey.}
\label{fig:stellarity}
\end{figure}

The correction works well for most of the stars, although we exclude
from the catalogs the immediate neighborhood. In the case of very
bright stars ( Fig~\ref{fig:removestar13}), multiple reflections between
the CCD and the optics make difficult the subraction of a median radial
profile and a faint halo is still visible after the correction.

\subsection{Star/galaxy separation}

SExtractor uses a neural network to separate star-like from extended
sources returning a stellarity index ({\sl CLASS\_STAR}) with values
between 1 (a perfect star-like object) and 0.  The distribution of
this index as a function of the magnitude in one of our fields (FLS18)
is shown in Figure~\ref{fig:stellarity}.

The standard neural network of SExtractor has been trained for seeing
FWHM values between 0.02 and 5.5 arcsec and for images that have 1.5
$<$ FWHM $<$ 5 pixels. It is therefore perfectly suited for our images.

Although at bright magnitudes two sequences can be easily
distinguished (see Figure~\ref{fig:stellarity}), for fainter magnitudes it becomes more
difficult to separate extended from point objects.
In order to select stars in an efficient way, we followed the technique
of Groenewegen et al. (2002) choosing a threshold which is a function
of the magnitude.
We consider that our objects are stars if:
\begin{eqnarray}
CLASS\_STAR > 0.85 & \qquad \textrm{ ... for } R < 23 \\
CLASS\_STAR > 0.90 & \qquad \textrm{ ... elsewhere}.
\end{eqnarray}

A drawback of this technique is that at faint magnitudes QSOs are also
classified as stars. However, using multi-band catalogs one can 
address the issue of separating stars and QSOs on the basis of their spectral
energy distributions.

\subsection{Source lists}

As an illustration, the tabulation of the first 30 entries in the FLS\_2 source catalog is presented
in Table~\ref{tbl:catalog}. All magnitudes are given in the Vega system. The Table lists:
\begin{itemize}
\item{Column 1:} the full IAU designation of the source;
\item{Columns 2-3:} right ascension and declination (J2000);
\item{Columns 4-7:} aperture (3 arcsec diameter) and total magnitudes and respective errors.
The total magnitude corresponds to the {\it MAG\_AUTO} magnitude measured by SExtractor.
The magnitudes have been not corrected for Galactic extinction.
The errors are those estimated by SExtractor and include only the shot-noise of the measured source and
background counts. Only objects detected with signal to noise S/N $\ge$ 3 (based on the total magnitude
errors) and without saturated pixels are included.
\item{Column 8:} an estimate of the S/N of the detection, from the errors estimated for
the total magnitude;
\item{Column 9:} the stellarity index computed by SExtractor;
\item{Column 10:} the Galactic extinction taken from Schlegel et al. (1998).
\end{itemize}

\begin{figure}[ht!]
   \centerline{\includegraphics[width=0.5\textwidth,angle=0]{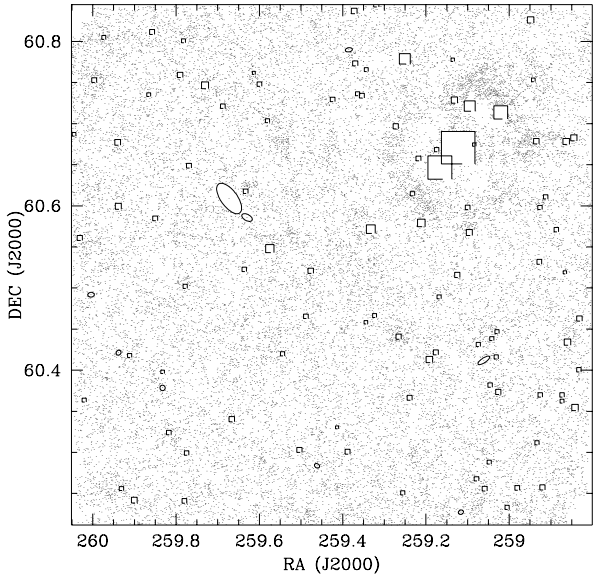}}
\caption{Projected distribution of galaxies extracted from the FLS\_6
image.  Square masks around the bright saturated
stars and  elliptical masks around bright extended galaxies in the
field are shown.}
\label{fig:fls_6}
\end{figure}
All the catalogs in the archive are in ASCII format.
Each catalog has been verified with several tests to check the reliability of the measured quantities
with other sets of data. The results of some of these tests are described in the next section.

\section{Survey performance}
The characteristics of the data obtained by the present survey are
summarized in Table~\ref{tbl:summary} which lists: in Col. 1 the name of the subfield,
in Cols. 2-3 the center of the subfield, in Col. 4 the seeing of the
combined image, in Cols. 5-6 the 3$\sigma$ and 5$\sigma$ limiting R Cousin magnitudes,
in Cols. 7-8  the number of galaxies and stars with S/N $\ge$ 3$\sigma$ which
are not saturated.

We measured the limiting R Cousin magnitude inside the aperture with
the highest signal-to-noise ratio in case of a Gaussian PSF dominated
by the sky. Considering a Gaussian profile, the S/N ratio
inside an aperture $R$ can be written as:
\begin{equation}
S/N=\frac{\int^{R}_{0} C e^{-2 \ln 2 r/W} dr}{\sqrt{\pi \frac{R^2}{\Delta^2} \sigma_S^2}},
\end{equation}
with $W$, FWHM, $C$, central intensity of the source, $\Delta$, pixel
size, and $\sigma_S$, the sky noise.  The $R$ for which S/N is maximal
corresponds to the $R$ for which $\partial (S/N) / \partial R = 0$.
This condition is realized at $R\approx 1.35 W/2$, i.e. an aperture
of 1.35~$\times$~FHWM.

Since we are considering in our catalogs an aperture of 3 arcsec and
our typical FHWM is 1.0 arcsec, these values are slightly deeper than
what one can find in our catalogs.

The fraction of spurious objects was estimated by creating catalogs
from the survey images multiplied by -1. Since ideally the noise is symmetric,
we can use these images to produce a catalog of spurious
sources by applying the same criteria of extraction which have been
used with the real images. Analyzing the central square degree,
false-positive detections occur only at faint magnitudes (R$>$23.8).
Considering all the 3$\sigma$ objects, in the magnitude interval 23.5$<$R$<$24.5
there are 150 false-positive detection per sq. degree corresponding to
0.5\% of the total number of sources. In the magnitude interval 24.5$<$R$<$25.5,
there are 119 false-positive detections per sq. degree which correspond
to 17\% of the total of number of sources detected in this magnitude range.

\subsection{Astrometry}

\begin{figure}[b!]
\centerline{\includegraphics[width=0.5\textwidth,angle=0]{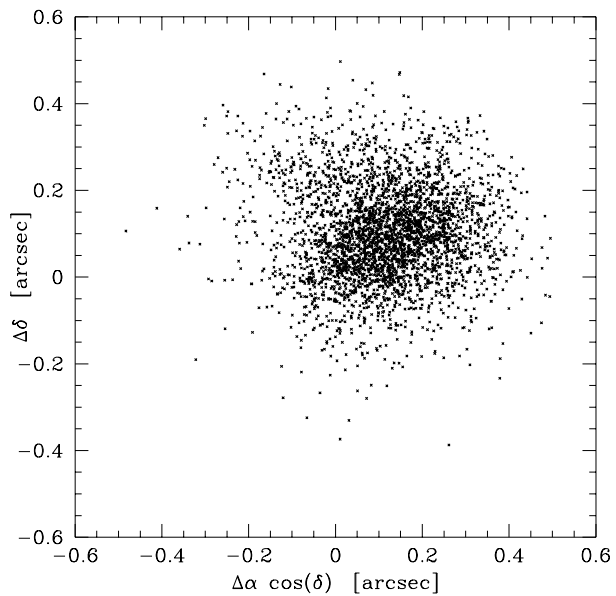}}
\caption{Comparison between the positions of stars in common with the
Sloan survey for the field FLS-18. Offsets are computed as our minus Sloan positions.
}
\label{fig:cfr_astro}
\end{figure}

To assess the accuracy of the astrometric calibration, we compared the 
positions of the stars inside the FLS field with those available
from the Sloan survey (Stoughton et al. 2002, Data Release 1) and with radio sources from 
the VLA survey in the FLS field (Condon et al. 2003). 

\begin{figure}[ht!]
 \centerline{\includegraphics[width=0.5\textwidth,angle=0]{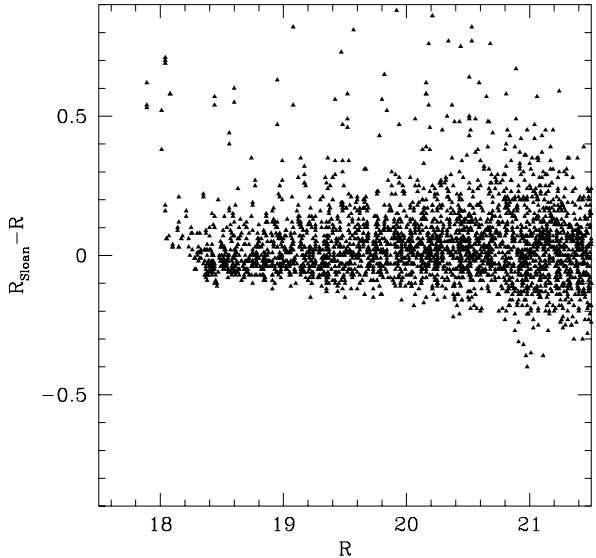}}
\caption{Comparison between the R magnitudes of stars in common with the
 Sloan Survey  in the field FLS-18 as a function of the R magnitude (Vega system).
}\label{fig:magoff18}
\end{figure}

In the comparison with the Sloan sources, we considered only good
objects (according to the flags) classified as stars in the Sloan
catalog with R magnitude between 18 and 21. Table~\ref{tbl:astrometry}
reports the number of stars used in the comparison and the offsets
between our and Sloan positions. Typical offsets are 0.1 arcsec in
right ascension and declination with an rms of 0.1 arcsec.  The
offsets between our catalog and the Sloan stars in the case of the
field FLS-18 are shown in Figure~\ref{fig:cfr_astro}. The comparison
with the VLA sources has been made considering all the optical
counterparts down to R=24 of non-extended radio sources.  The offsets
from the VLA positions are also of order 0.1 arcsec.

\begin{figure}[ht!]
\centerline{\includegraphics[width=0.5\textwidth,angle=0]{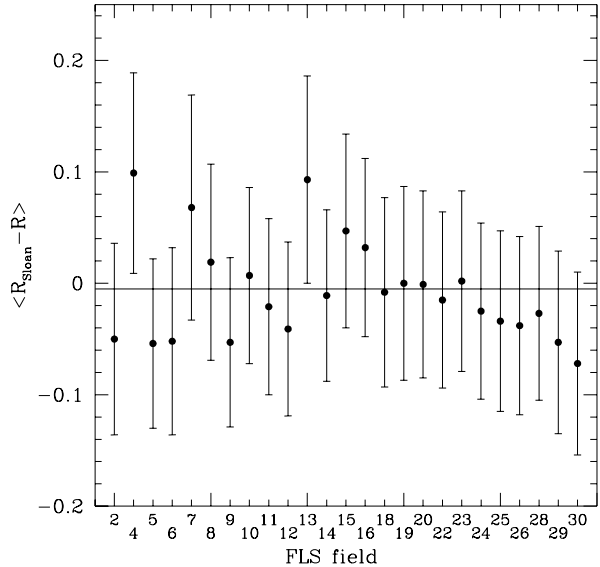}}
\caption{Offsets between our and Sloan magnitudes for stars with $18<R<21$ 
in the various FLS fields. The average difference is $-0.007 \pm 0.016$. }
\label{fig:magoff}
\end{figure}

The systematic offset between ours and Sloan positions comes from the
fact that our astrometry is based on the USNO-A2 catalog, while Sloan
takes Tycho2 stars as reference. Comparing the Sloan and USNO-A2
positions in the whole FLS field, we have an offset of $\Delta\alpha=0.13\pm0.28$
and $\Delta\delta=0.15\pm0.34$ arcsec which is in complete agreement with the
offsets found in Table~\ref{tbl:astrometry}. 

Finally, we compared our R-band catalog to the GSC~II finding an
offset of $\Delta\alpha=-0.05\pm0.18$ and $\Delta\delta=0.22\pm0.15$ arcsec. 
While not large, this offset should be noted when making comparison to
data sets that used the GSC~II for reference.

Nevertheless, the remarkably small rms in both directions obtained comparing our
and Sloan positions suggests an intrinsic accuracy of $\lesssim 0.1$
arcsec for each catalog which is well within the
requirements for slit/fiber positioning, an essential requirement for
public surveys.

\subsection{Photometry}

The Sloan survey also allows us to compare the photometric calibration.
Although the magnitude system used in this survey is different from ours,
we can obtain a relationship between the R Cousin magnitude and the 
Sloan magnitudes using the 92 Landolt stars which have been observed
by the Sloan group to calibrate their observations (Smith et al. 2002).
Before making the comparison, we have converted the r' and i' magnitudes to the SDSS
2.5m natural system, using the equations:
\begin{eqnarray}
r_{SDSS}&=&r'+0.035 \ (r'-i'-0.21),\\
i_{SDSS}&=&i'+0.041 \ (r'-i'-0.21),
\label{eqt:sdssphot}
\end{eqnarray}
as explained at the Sloan web
site\footnote{www.sdss.org/DR1/algorithms/jeg\_photometric\_eq\_dr1.html}.
Then, we have obtained a relationship between R and the Sloan colors
with a least square fit:
\begin{equation}
R = -0.16 + r_{SDSS} - 0.26 \ (r_{SDSS} - i_{SDSS}).
\label{eqt:landolt_sloan}
\end{equation}
Using a biweight estimator, the difference between the real R and the
value estimated with the relationship (\ref{eqt:landolt_sloan}) is on
average of 0.0001 (with an rms of 0.007).  In spite of the large
scatter, the relationship is useful from a statistical point of view
since we are interested only in confirming our magnitude zero-point.

We have therefore compared the magnitudes of the stars in the FLS
fields with the values deduced from the Sloan survey with the equation
(\ref{eqt:landolt_sloan}).  Table~\ref{tbl:astrometry} summarizes the
median differences in magnitude between our measurements ({\it
auto-magnitudes}) and the {\it model} complete magnitude as computed
by Sloan for stars with magnitude $18<R<21$ in the different FLS
fields. In Figure~\ref{fig:magoff18} we show the distribution of the
magnitude offsets in the case of the central field FLS-18, while
in Figure~\ref{fig:magoff} we illustrate the offsets for the various
fields. Our calibration agrees on average with the Sloan one, since
the average difference between our and Sloan measurements
is  $-0.007 \pm 0.016$. The biggest differences are found in the external
subfields where the relative zero is not well constrained due to the low
number of stars in common between adjacent subfields (see Figure~\ref{fig:magoff}).

\begin{figure}[ht!]
\centerline{\includegraphics[width=0.5\textwidth,angle=0]{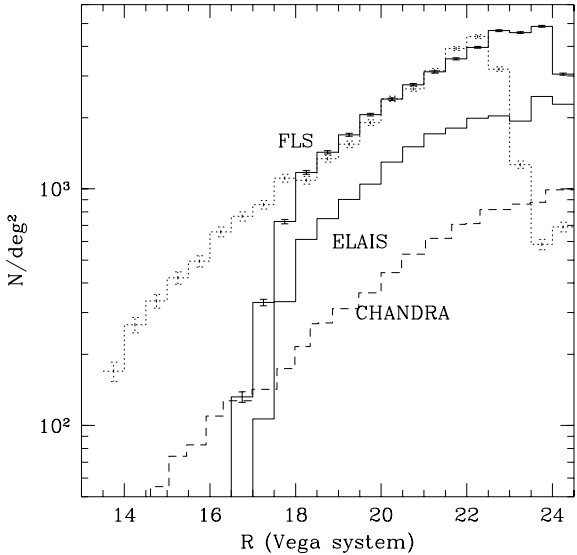}}
\caption{Star counts in the central square degree of the FLS region
from our R images (solid line) and Sloan Digital Sky Survey catalogs
(dotted line).  For comparison, the dashed line refers to the counts in
the Chandra South region (Groenewegen et al., 2002).  }
\label{fig:starcounts}
\end{figure}
\begin{figure}[ht!]
\centerline{\includegraphics[width=0.5\textwidth,angle=0]{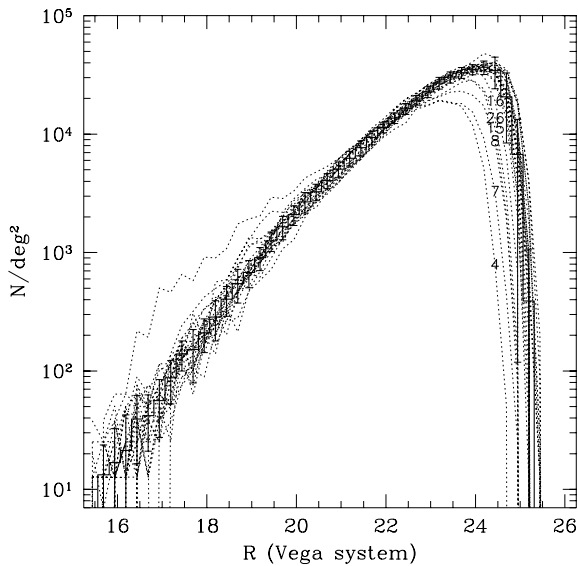}}
\caption{Galaxy counts in the various FLS subfields. The numbers refer
to the FLS subfields with shallowest depths.
}
\label{fig:fieldscounts}
\end{figure}

\subsection{Number counts}

Counting galaxies and stars as a function of the magnitude allows one
to evaluate the overall characteristics of a catalog as depth and
homogeneity.

\begin{figure}[!ht]
\centerline{\includegraphics[width=0.5\textwidth,angle=0]{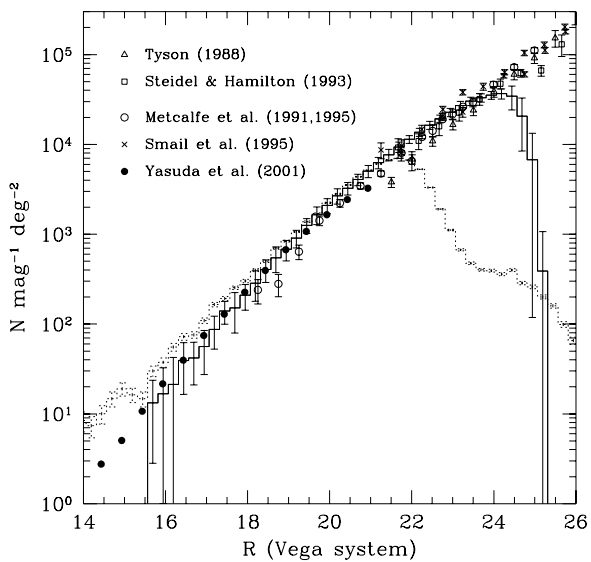}}
\caption{Galaxy counts in the FLS region
 from our R images (solid line)
and Sloan Digital Sky Survey catalogs (dotted line).
Results from counts in other sky regions are overplotted.
}
\label{fig:galscounts}
\end{figure}

In Figures~\ref{fig:starcounts} and~\ref{fig:galscounts} we show star and
galaxy counts in the FLS region and compare them with analogous counts
using the Sloan Digital Sky Survey (Stoughton et al. 2002).

To compare the two distributions, we have transformed the SDSS magnitudes
into the R Vega magnitudes using the relationship~\ref{eqt:landolt_sloan}.

In the case of star counts (Fig.~\ref{fig:starcounts}), the counts from our
survey and the SDSS agree very well between R=18 and R=22. For magnitudes
brighter than R=18, most of the stars detected in our survey are saturated
and do not appear in our catalogs.
Star counts drop very rapidly for magnitudes fainter than R=24
since the profile criterion used for the star/galaxy separation fails
for faint objects.
For comparison,  we show in Figure~\ref{fig:starcounts} the star counts in the
ELAIS field and those in the Chandra field (Groenewegen et al., 2002).
These fields, which lie at higher galactic latitudes are, as expected,
less populated by stars.

To evaluate the variation in the number counts due to the varying
observing conditions, we have computed the counts for each of the
subfield in the FLS field. In Figure~\ref{fig:fieldscounts} we show these
counts as well as the median counts with error bars corresponding to
the standard deviation as measured from the observed scatter in the
counts of the different subfields.  One can easily see that a few
subfields (\#4, \#7, \#8, \#15 and \#26) are less deep than the other
ones, as expected from the quantities measured in
Table~\ref{tbl:summary}. Fortunately, these fields are external and
have been only partially covered by Spitzer observations.  The other
20 fields in the FLS are quite homogeneous.

Median counts are then reported in Figure~\ref{fig:galscounts} to
compare them with the results from other surveys. The dotted line
corresponds the counts from SDSS in the FLS field (Stoughton et al. 2002)
computed transforming the SDSS magnitudes into the R Vega magnitudes
using the relationship in equation (\ref{eqt:landolt_sloan}).  
The points from the
general SDSS counts (Yasuda et al., 2001) have been approximatively
transformed using the relationship by Fukugita et al. (1995) assuming
that galaxies have $R-r'$ colors typical of spirals at redshift of
0.2~--~0.5.  At the faint end of the counts, results from several deep
surveys are reported.

\begin{figure}[!ht]
\centerline{\includegraphics[width=0.5\textwidth,angle=0]{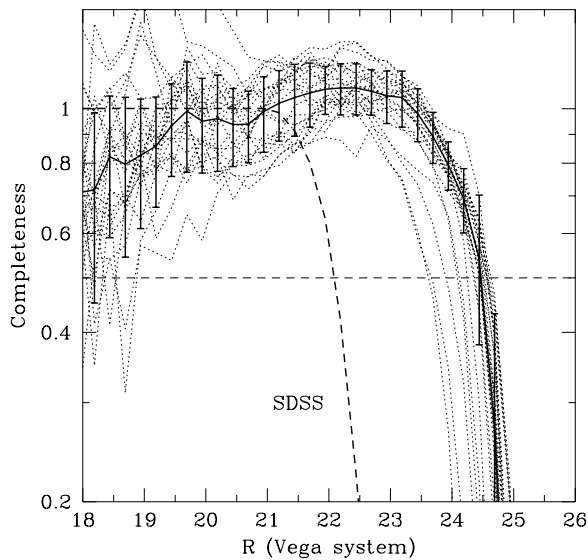}}
\caption{Ratio between the number galaxies in our FLS survey and that
of SDSS in the same region for $R<21$ and that of median counts from
several deep surveys at $R>21$. Dotted thin lines show the completeness
of single FLS subfields, while the dashed line shows the completeness
of the SDSS survey in the FLS field.
}
\label{fig:completeness}
\end{figure}

As an estimate of the completeness of our images we have compared our
counts to the SDSS counts in the FLS field for R$<$21 and to median
counts from the other deep surveys at magnitudes fainter
than $R=21$.  In Figure~\ref{fig:completeness} we show the completeness of
the single subfields in the FLS field (thin lines) and those of the
global field and the SDSS survey (continuous and dashed thick lines,
respectively).  Our survey is deeper than the SDSS data by almost three
magnitudes.  It is 50\% complete around R=24.5. This estimate is
conservative, since the number of spurious detection
at R=24.5 is still relatively small (less than 0.5\%).

Finally, for magnitudes brighter than R=18, the galaxy catalogs are
slightly incomplete since a few extended objects are saturated.


\section{Summary}
A deep NOAO/KPNO survey in the R band has been carried out to observe a
field of more than 9 square degrees centered at 17:18:00 +59:30:00
(J2000) aimed to find optical counterparts for the {\sl First Look
Survey} which surveys 7 different infrared wavelengths
with the instruments IRAC and MIPS using the Spitzer Space Telescope. 
Another 2.3 square degrees have been surveyed in the
ELAIS-N1 region which will be observed in the \sst\ SWIRE Legacy 
survey.  This paper describes the observation strategy, the data
reduction and the products which are publicly available to the
astronomical community on the World Wide Web at
the Spitzer Science Center and the NOAO Science Archives.

The overall quality of the data is good and homogeneous: the average
seeing is 1.1 and typically varies between 0.9 and 1.2. The limiting
magnitude of the images, measured inside an aperture of 1.35 $\times$
FWHM for which the S/N ratio is maximal, is around
R=25.5 at 5$\sigma$, deep enough to detect optical counterparts for a
substantial fraction of the new {\it Spitzer} selected objects.

An average number of  35000 extragalactic sources are detected in each 
subfield (40' $\times$ 40', approximatively) with a 50\% completeness limit
of R=24.5  as deduced by comparing the counts with other deeper surveys.

Images and catalogs are available to the astronomical community along
with the first release of the FLS infrared data to exploit in the
best way the wealth of extragalactic data expected from the new
infrared observatory {\it Spitzer}.

\acknowledgments

We are grateful to those that built, maintain, and operate the
MOSAIC-1 camera and Mayall 4m telescope at Kitt Peak National
Observatory, a part of the National Optical Astronomy Observatory,
which is operated by the Association of Universities for Research in
Astronomy, Inc. (AURA), under a cooperative agreement with the
National Science Foundation.  We thank the former director of NOAO,
Dr. Sidney Wolff, for allocating the directors discretionary time used
to gather the optical data we present in this paper, M. J. I. Brown
for assistance with software used to prepare the images for the data
release, and T. Lauer for useful discussions about photometric depth
measurements. We wish to thank also the anonymous referee for his
comments about the Sloan photometric calibration. We acknowledge the
support of the Extragalactic First Look Survey Team members at the
Spitzer Science Center (B.T. Soifer, P. Appleton, L. Armus,
S. Chapman, P. Choi, F. Fang, D. Frayer, I. Heinrichsen, G. Helou,
M. Im., M. Lacy, S. Laine, F. Marleau, D. Shupe, G. Squires J. Surace,
H. Teplitz, G. Wilson, L. Yan, J. Colbert, and I. Drozdovsky)


\newpage

\begin{deluxetable}{ccccc}
\tablecolumns{5}
\tablewidth{0pt}
\tabletypesize{\scriptsize}
\tablecaption{Main features of the filter used in the MOSAIC-1 observations.\label{tbl:filter}}
\tablehead{
\colhead{Filter} & \colhead{KPNO ID}   & \colhead{$\lambda_{eff}$}   &
\colhead{FWHM} & \colhead{Peak Throughput} \\
&&\colhead{{\rm \,\AA}}&\colhead{{\rm \,\AA}}&
}
\startdata
R & R Harris k1004 & 6440 & 1510 & 86.2\% \\
\enddata
\end{deluxetable}

\begin{deluxetable}{cccc}
\tablecolumns{5}
\tablewidth{0pt}
\tabletypesize{\scriptsize}
\tablecaption{Nominal coordinates of the observed subfields.\label{tbl:coordinates}}
\tablehead{
\colhead{ID} & \colhead{$\alpha$ (J2000)}   & \colhead{$\delta$
(J2000)} & Nr. of 10 minute obs.
}
\startdata
FLS~~2  & 17:17:48.12 & +61:06:00.0 & 3\\
FLS~~4  & 17:08:32.33 & +60:32:00.0 & 3\\
FLS~~5  & 17:13:13.17 & +60:32:00.0 & 3\\
FLS~~6  & 17:17:45.01 & +60:32:00.0 & 3\\
FLS~~7  & 17:22:25.85 & +60:32:00.0 & 3\\
FLS~~8  & 17:27:06.69 & +60:32:00.0 & 3\\
FLS~~9  & 17:05:44.94 & +59:57:00.0 & 3\\
FLS~10 & 17:10:12.89 & +59:57:00.0 & 5\\
FLS 11 & 17:14:48.65 & +59:57:00.0 & 5\\
FLS 12 & 17:19:24.61 & +59:57:00.0 & 4\\
FLS 13 & 17:23:59.47 & +59:57:00.0 & 5\\
FLS 14 & 17:28:29.33 & +59:57:00.0 & 3\\
FLS 15 & 17:05:56.30 & +59:22:00.0 & 3\\
FLS 16 & 17:10:25.14 & +59:22:00.0 & 3\\
FLS 17 & 17:14:54.99 & +59:22:00.0 & 4\\
FLS 18 & 17:19:20.82 & +59:22:00.0 & 4\\
FLS 19 & 17:23:53.67 & +59:22:00.0 & 3\\
FLS 20 & 17:28:22.51 & +59:22:00.0 & 3\\
FLS 22 & 17:10:18.66 & +58:47:00.0 & 3\\
FLS 23 & 17:14:46.17 & +58:47:00.0 & 3\\
FLS 24 & 17:19:12.00 & +58:47:00.0 & 3\\
FLS 25 & 17:23:37.32 & +58:47:00.0 & 3\\
FLS 26 & 17:28:02.83 & +58:47:00.0 & 3\\
FLS 28 & 17:12:58.88 & +58:12:30.0 & 3\\
FLS 29 & 17:17:19.44 & +58:12:00.0 & 3\\
FLS 30 & 17:21:40.00 & +58:12:30.0 & 3\\
\hline
ELAIS~~2 & 16:10:01.00 & +55:23:06.0 & 3\\
ELAIS~~5 & 16:08:00.00 & +54:48:06.0 & 5\\
ELAIS~~6 &16:12:06.76 &  +54:48:06.0 & 7\\ 
ELAIS~~9 &16:08:02.04 & +54:13:06.2 & 9\\
ELAIS~10 &16:12:06.76 & +54:13:06.2 & 4\\
\enddata
\end{deluxetable}


\begin{deluxetable}{ccccc}
\tablecolumns{5}
\tablewidth{0pt}
\tabletypesize{\scriptsize}
\tablecaption{Log of observations.\label{tbl:log}}
\tablehead{
\colhead{Date} & \colhead{Subfield} & \colhead{Exposure}  & \colhead{Seeing} & \colhead{Photometric}\\ 
\colhead{} & \colhead{Number} & \colhead{Time}  & \colhead{Range} & \colhead{Conditions}\\
\colhead{} & \colhead{} & \colhead{(s)}  & \colhead{(arcsec)} & \colhead{} 
}
\startdata
2000 May 4 & FLS04 & 3 $\times$ 600. & 1.4-1.5  & light cirri on sunset\\
           & FLS05 & 3 $\times$ 600. & 1.4-1.45 & photometric\\
           & FLS06 & 3 $\times$ 600. & 1.5-1.6  & \\
           & FLS07 & 3 $\times$ 600. & 1.2-1.5  & \\
           & FLS08 & 3 $\times$ 600. & 1.2-1.3  & \\
           & FLS10 & 5 $\times$ 600. &0.98-1.15 & \\
           & FLS11 & 3 $\times$ 600. &0.97-1.15 & \\
           & FLS12 & 4 $\times$ 600. &0.79-1.3  & \\
           & FLS17 & 4 $\times$ 600. &0.8-0.95  & \\
2000 May 5 & ELAIS05 & 3 $\times$ 600. & 1.07-1.25& photometric\\
           & FLS15 & 3 $\times$ 600. &1.05-1.12 & \\
           & FLS16 & 3 $\times$ 600. &0.98-1.04 & \\
           & FLS20 & 3 $\times$ 600. &0.88-0.9  & \\
           & FLS19 & 3 $\times$ 600. &0.85-0.87 & \\
           & FLS18 & 4 $\times$ 600. &0.77-0.88 & \\
           & FLS24 & 3 $\times$ 600. &0.87-0.95 & \\
           & FLS23 & 3 $\times$ 600. &0.9-0.94  & \\
           & FLS22 & 3 $\times$ 600. &0.81-0.86 & \\
           & FLS25 & 3 $\times$ 600. &0.85-0.9  & \\
           & FLS13 & 1 $\times$ 600. &0.9       & \\
2000 May 6 & FLS13 & 4 $\times$ 600. &1.05-1.25 & light cirri\\
           & FLS11* & 2 $\times$ 600. &0.95-1.1 & non-photometric\\
           & ELAIS06* & 4 $\times$ 600. &0.88-0.95&\\
           & ELAIS09* & 4 $\times$ 600. &0.86-0.95&\\
           & ELAIS10* & 4 $\times$ 600. &0.87-1.0&\\
           & FLS14* & 4 $\times$ 600. &0.84-0.85& \\
2000 May 7 & ELAIS02 & 3 $\times$ 600. &1.09-1.2& cirri\\
           & FLS26 & 3 $\times$ 600. &1.0-1.2   & non-photometric\\
           & FLS28 & 3 $\times$ 600. &0.96-1.0  & \\
           & FLS29 & 3 $\times$ 600. &0.95-1.2  & \\
           & FLS30 & 3 $\times$ 600. &1.02-1.07 & \\
           & FLS09 & 3 $\times$ 600. &1.05-1.1  & \\
           & FLS02 & 3 $\times$ 600. &0.94-0.98 & \\
           & FLS05 & 3 $\times$ 600. &0.9-1.06  & \\
           & FLS06 & 3 $\times$ 600. &0.9-0.91  & \\
           & ELAIS05 & 2 $\times$ 600. &0.9-1.1 & \\
           & ELAIS06 & 3 $\times$ 600. &0.9-1.1 & \\
           & FLS14 & 3 $\times$ 600. &0.86-0.92 & \\
2000 May 9 & ELAIS09 & 5 $\times$ 600. &1.0-1.2 & non-photometric\\
\enddata
\tablenotetext{*}{A subfield observed 24' south of the originally intended position.}
\end{deluxetable}

 \clearpage

\begin{deluxetable}{cccccccrcc}
\tablecolumns{10}
\tablewidth{0pt}
\tabletypesize{\scriptsize}
\tablecaption{First 30 entries of the FLS\_2 source list.\label{tbl:catalog}}
\tablehead{
\colhead{Identification} & \colhead{$\alpha$ (J2000) } & \colhead{$\delta$ (J2000)} 
 & \colhead{$m_{aper}$} & \colhead{$\epsilon$}& \colhead{$m_{tot}$}& \colhead{$\epsilon$} &
 \colhead{S/N}  & \colhead{Class}  & \colhead{Ext} \\
\colhead{(1)} &\colhead{(2)} &\colhead{(3)} &\colhead{(4)} &\colhead{(5)} &
\colhead{(6)} &\colhead{(7)} &\colhead{(8)} &\colhead{(9)} &\colhead{(10)}
}
\startdata
FLS\_R\_J171814.3+604643& 17:18:14.303& +60:46:43.93& 21.44&  0.02& 21.44&  0.03&  38.8& 0.997& 0.071\\
FLS\_R\_J171804.5+604644& 17:18:04.511& +60:46:44.00& 23.60&  0.12& 23.66&  0.12&   8.8& 0.753& 0.071\\
FLS\_R\_J171847.4+604644& 17:18:47.496& +60:46:44.22& 24.12&  0.20& 24.19&  0.21&   5.2& 0.735& 0.073\\
FLS\_R\_J171822.5+604644& 17:18:22.536& +60:46:44.07& 22.99&  0.07& 22.95&  0.09&  12.2& 0.975& 0.071\\
FLS\_R\_J171840.3+604644& 17:18:40.368& +60:46:44.40& 23.61&  0.12& 23.60&  0.12&   9.1& 0.936& 0.072\\
FLS\_R\_J171729.6+604645& 17:17:29.639& +60:46:45.47& 24.08&  0.19& 23.53&  0.18&   6.0& 0.784& 0.069\\
FLS\_R\_J171739.6+604645& 17:17:39.672& +60:46:45.69& 24.65&  0.32& 24.36&  0.17&   6.4& 0.802& 0.070\\
FLS\_R\_J171808.7+604643& 17:18:08.784& +60:46:43.96& 20.72&  0.01& 20.69&  0.01& 104.4& 0.983& 0.071\\
FLS\_R\_J171824.5+604644& 17:18:24.575& +60:46:44.11& 20.84&  0.01& 20.73&  0.02&  67.4& 0.983& 0.071\\
FLS\_R\_J171817.5+604644& 17:18:17.567& +60:46:44.65& 22.51&  0.04& 22.49&  0.06&  18.8& 0.960& 0.071\\
FLS\_R\_J171851.9+604644& 17:18:51.984& +60:46:44.61& 20.75&  0.01& 20.71&  0.01&  94.4& 0.984& 0.073\\
FLS\_R\_J171914.2+604645& 17:19:14.232& +60:46:45.44& 22.62&  0.05& 22.58&  0.06&  18.3& 0.973& 0.074\\
FLS\_R\_J171734.8+604646& 17:17:34.872& +60:46:46.05& 23.22&  0.09& 23.12&  0.09&  12.0& 0.851& 0.070\\
FLS\_R\_J171912.1+604647& 17:19:12.191& +60:46:47.13& 24.28&  0.23& 24.29&  0.23&   4.7& 0.652& 0.074\\
FLS\_R\_J171841.0+604646& 17:18:41.087& +60:46:46.91& 23.66&  0.13& 23.64&  0.13&   8.6& 0.944& 0.072\\
FLS\_R\_J171747.7+604648& 17:17:47.712& +60:46:48.53& 24.61&  0.31& 24.45&  0.18&   6.1& 0.758& 0.070\\
FLS\_R\_J171726.8+604646& 17:17:26.807& +60:46:46.12& 23.29&  0.09& 22.58&  0.11&  10.3& 0.851& 0.069\\
FLS\_R\_J171827.8+604646& 17:18:27.887& +60:46:46.99& 23.30&  0.09& 22.86&  0.11&   9.6& 0.873& 0.072\\
FLS\_R\_J171834.5+604648& 17:18:34.560& +60:46:48.28& 23.94&  0.17& 23.89&  0.14&   7.6& 0.945& 0.072\\
FLS\_R\_J171838.2+604644& 17:18:38.279& +60:46:44.40& 20.22&  0.01& 20.19&  0.01& 155.1& 0.985& 0.072\\
FLS\_R\_J171631.4+604647& 17:16:31.487& +60:46:47.20& 23.47&  0.11& 21.60&  0.06&  18.8& 0.947& 0.064\\
FLS\_R\_J171731.9+604646& 17:17:31.967& +60:46:46.99& 23.09&  0.08& 22.29&  0.09&  11.6& 0.749& 0.069\\
FLS\_R\_J171821.9+604645& 17:18:21.911& +60:46:45.91& 23.17&  0.08& 22.98&  0.11&   9.6& 0.434& 0.071\\
FLS\_R\_J171719.7+604649& 17:17:19.704& +60:46:49.00& 23.81&  0.15& 23.74&  0.21&   5.1& 0.928& 0.068\\
FLS\_R\_J171618.3+604647& 17:16:18.312& +60:46:47.63& 24.48&  0.27& 24.22&  0.18&   6.0& 0.671& 0.063\\
FLS\_R\_J171843.1+604648& 17:18:43.128& +60:46:48.39& 23.74&  0.14& 23.84&  0.14&   7.6& 0.955& 0.073\\
FLS\_R\_J171733.5+604649& 17:17:33.503& +60:46:49.40& 23.62&  0.12& 23.33&  0.16&   6.7& 0.974& 0.069\\
FLS\_R\_J171730.6+604649& 17:17:30.671& +60:46:49.58& 24.10&  0.19& 23.08&  0.14&   7.8& 0.879& 0.069\\
FLS\_R\_J171851.1+604645& 17:18:51.143& +60:46:45.26& 19.53&  0.00& 19.49&  0.00& 258.5& 0.985& 0.073\\
FLS\_R\_J171646.7+604647& 17:16:46.751& +60:46:47.28& 22.53&  0.05& 22.05&  0.07&  16.4& 0.043& 0.065\\
\enddata
\end{deluxetable}
 \clearpage

\begin{deluxetable}{cccccccc}
\tablecolumns{8}
\tabletypesize{\scriptsize}
\tablecaption{Properties of images and extracted catalogs\label{tbl:summary}}
\tablewidth{0pt}
\tablehead{
\colhead{ID} & \colhead{RA} & \colhead{DEC}  & \colhead{Seeing}   & \colhead{$m_{lim}(5\sigma)$}& \colhead{$m_{lim}(3\sigma)$}& \colhead{$N_{gals}$}   & \colhead{$N_{stars}$}\\
 & \colhead{(J2000)} & \colhead{(J2000)}  & \colhead{(arcsec)}   & \colhead{(mag)} & \colhead{(mag)}&&
}
\startdata
FLS 2     & 17:17:45.70 & 61:05:44.12 & 1.08 & 25.59 & 26.15 &34766& 6787\\
FLS 4     & 17:08:27.23 & 60:32:10.74 & 1.57 & 24.72 & 25.28 &18016& 2962\\
FLS 5     & 17:13:10.74 & 60:31:44.01 & 1.06 & 25.64 & 26.20 &36372& 5012\\
FLS 6     & 17:17:42.13 & 60:31:42.33 & 1.02 & 25.78 & 26.34 &36988& 5050\\
FLS 7     & 17:22:21.27 & 60:32:06.70 & 1.54 & 24.86 & 25.42 &19959& 1316\\
FLS 8     & 17:27:02.18 & 60:32:07.10 & 1.36 & 25.09 & 25.64 &22392& 3049\\
FLS 9     & 17:05:42.63 & 59:56:46.44 & 1.14 & 25.58 & 26.13 &34985& 5254\\
FLS 10    & 17:10:07.30 & 59:57:07.22 & 1.24 & 25.63 & 26.19 &32253& 3309\\
FLS 11-17 & 17:14:47.10 & 59:39:33.96 & 1.04 & 25.86 & 26.41 &72709& 6865\\
FLS 12    & 17:19:19.96 & 59:57:03.33 & 1.00 & 25.86 & 26.42 &35905& 4352\\
FLS 13    & 17:23:54.12 & 59:57:09.50 & 1.16 & 25.41 & 25.96 &30797& 5496\\
FLS 14    & 17:28:26.67 & 59:56:39.57 & 0.98 & 25.51 & 26.06 &34672& 5114\\
FLS 15    & 17:05:50.01 & 59:22:06.83 & 1.22 & 25.14 & 25.69 &25218& 4450\\
FLS 16    & 17:10:19.12 & 59:22:07.62 & 1.17 & 25.25 & 25.80 &26996& 4313\\
FLS 18    & 17:19:14.84 & 59:22:07.54 & 0.92 & 25.91 & 26.46 &37922& 4228\\
FLS 19    & 17:23:48.01 & 59:22:07.27 & 0.97 & 25.64 & 26.19 &34771& 4409\\
FLS 20    & 17:28:17.04 & 59:22:08.13 & 1.00 & 25.59 & 26.15 &35297& 4278\\
FLS 22    & 17:10:12.41 & 58:47:07.63 & 0.94 & 25.77 & 26.33 &35201& 4259\\
FLS 23    & 17:14:40.18 & 58:47:07.50 & 1.06 & 25.67 & 26.22 &34624& 4313\\
FLS 24    & 17:19:06.04 & 58:47:04.74 & 1.03 & 25.71 & 26.26 &31151& 4277\\
FLS 25    & 17:23:30.85 & 58:47:10.81 & 0.99 & 25.76 & 26.31 &33847& 4682\\
FLS 26    & 17:28:00.70 & 58:46:43.81 & 1.22 & 25.21 & 25.77 &22636& 5538\\
FLS 28    & 17:12:56.31 & 58:12:15.54 & 1.12 & 25.37 & 25.93 &30609& 5830\\
FLS 29    & 17:17:16.96 & 58:11:44.23 & 1.12 & 25.52 & 26.07 &34524& 6264\\
FLS 30    & 17:21:37.91 & 58:12:14.09 & 1.20 & 25.49 & 26.05 &38493&10099\\
\hline			 	       	      	    	                 
ELAIS 2   & 16:09:58.00 & 55:22:51.47 & 1.20 & 25.18 & 25.74 &26960& 3802\\
ELAIS 5   & 16:07:55.71 & 54:47:58.39 & 1.22 & 25.59 & 26.14 &33876& 3240\\
ELAIS 6   & 16:11:59.38 & 54:47:43.34 & 1.11 & 25.49 & 26.05 &34736& 3406\\
ELAIS 6S  & 16:11:58.89 & 54:23:29.02 & 0.98 & 25.82 & 26.38 &41506& 3160\\
ELAIS 9   & 16:07:59.66 & 54:12:59.29 & 1.11 & 25.77 & 26.32 &38231& 2926\\
ELAIS 9S  & 16:07:58.67 & 53:48:27.43 & 1.01 & 25.78 & 26.33 &38167& 3538\\
ELAIS 10S & 16:11:56.45 & 53:48:23.99 & 1.02 & 25.72 & 26.28 &36497& 3496\\
\enddata
\end{deluxetable}

 \clearpage

\begin{deluxetable}{ccrrrcrr}
\tablecolumns{8}
\tabletypesize{\scriptsize}
\tablecaption{Astrometry and photometry: comparison with SDSS and VLA.\label{tbl:astrometry}}
\tablewidth{0pt}
\tablehead{
\colhead{ID} & \colhead{Nr. of stars} & \colhead{$< R_{Sloan}-R>$}  & 
\colhead{$\alpha-\alpha_{Sloan}$}   & \colhead{$\delta-\delta_{Sloan}$}&
 \colhead{Nr. of VLA}& \colhead{$\alpha-\alpha_{VLA}$}   & \colhead{$\delta-\delta_{VLA}$}\\
 &  & \colhead{mag}  & \colhead{arcsec}   & \colhead{arcsec} & 
\colhead{sources}& \colhead{arcsec}   & \colhead{arcsec}
}

\startdata
FLS 2    &2004&-0.050$\pm$0.086& 0.12$\pm$0.12& 0.07$\pm$0.09& 23& 0.11$\pm$0.44& 0.04$\pm$0.38\\
FLS 4    &1523& 0.100$\pm$0.090& 0.19$\pm$0.12&-0.01$\pm$0.10&   &              &              \\
FLS 5    &1523&-0.054$\pm$0.076& 0.12$\pm$0.12& 0.07$\pm$0.09& 84& 0.09$\pm$0.34& 0.24$\pm$0.35\\
FLS 6    &1967&-0.052$\pm$0.084& 0.17$\pm$0.13& 0.15$\pm$0.10&117& 0.17$\pm$0.37& 0.28$\pm$0.36\\
FLS 7    &1501& 0.068$\pm$0.101& 0.11$\pm$0.13& 0.11$\pm$0.10& 76& 0.10$\pm$0.39& 0.18$\pm$0.31\\
FLS 8    &1571& 0.019$\pm$0.088& 0.14$\pm$0.13& 0.18$\pm$0.09&   &              &              \\
FLS 9    &1319&-0.053$\pm$0.076& 0.21$\pm$0.11& 0.03$\pm$0.09&   &              &              \\
FLS 10   &1509& 0.007$\pm$0.079& 0.16$\pm$0.13& 0.09$\pm$0.09&125& 0.11$\pm$0.33& 0.09$\pm$0.39\\
FLS 11-17&2810&-0.021$\pm$0.079& 0.12$\pm$0.14& 0.13$\pm$0.08&269& 0.15$\pm$0.35& 0.08$\pm$0.38\\
FLS 12   &2132&-0.041$\pm$0.079& 0.12$\pm$0.12& 0.18$\pm$0.08&121& 0.27$\pm$0.32& 0.22$\pm$0.29\\
FLS 13   &1412& 0.093$\pm$0.093& 0.09$\pm$0.14& 0.15$\pm$0.10&128& 0.08$\pm$0.37& 0.12$\pm$0.37\\
FLS 14   &1590&-0.011$\pm$0.078& 0.08$\pm$0.13& 0.10$\pm$0.09& 29& 0.07$\pm$0.35& 0.13$\pm$0.47\\
FLS 15   &1335& 0.047$\pm$0.087& 0.22$\pm$0.12& 0.07$\pm$0.11&   &              &              \\
FLS 16   &1379& 0.032$\pm$0.080& 0.18$\pm$0.14& 0.09$\pm$0.09&106& 0.21$\pm$0.40& 0.12$\pm$0.35\\
FLS 18   &2294&-0.008$\pm$0.085& 0.12$\pm$0.13& 0.09$\pm$0.10&144& 0.21$\pm$0.32& 0.12$\pm$0.30\\
FLS 19   &1614&-0.000$\pm$0.087& 0.11$\pm$0.12& 0.14$\pm$0.09&142& 0.19$\pm$0.37& 0.11$\pm$0.37\\
FLS 20   &1608&-0.001$\pm$0.084& 0.06$\pm$0.13& 0.11$\pm$0.10& 58& 0.02$\pm$0.31& 0.14$\pm$0.40\\
FLS 22   &1461&-0.015$\pm$0.079& 0.17$\pm$0.13&-0.11$\pm$0.10& 61& 0.30$\pm$0.35&-0.02$\pm$0.35\\
FLS 23   &1510& 0.002$\pm$0.081& 0.10$\pm$0.15& 0.03$\pm$0.09&121& 0.09$\pm$0.35& 0.07$\pm$0.37\\
FLS 24   &2044&-0.025$\pm$0.079& 0.03$\pm$0.13& 0.07$\pm$0.12&143& 0.05$\pm$0.38& 0.07$\pm$0.37\\
FLS 25   &2022&-0.034$\pm$0.082& 0.04$\pm$0.12& 0.05$\pm$0.09&117& 0.13$\pm$0.38& 0.08$\pm$0.36\\
FLS 26   &1539&-0.038$\pm$0.080& 0.05$\pm$0.13& 0.00$\pm$0.09& 34& 0.04$\pm$0.40& 0.17$\pm$0.40\\
FLS 28   &1273&-0.027$\pm$0.078&-0.11$\pm$0.16& 0.05$\pm$0.16&   &              &              \\
FLS 29   &1769&-0.053$\pm$0.082&-0.01$\pm$0.14& 0.22$\pm$0.08& 57& 0.02$\pm$0.41& 0.18$\pm$0.35\\
FLS 30   &3198&-0.072$\pm$0.082& 0.02$\pm$0.13& 0.17$\pm$0.11& 23& 0.09$\pm$0.40& 0.25$\pm$0.30\\
average  &    &-0.007$\pm$0.016& 0.10$\pm$0.03& 0.09$\pm$0.02&   & 0.12$\pm$0.08& 0.13$\pm$0.08\\
\enddata
\end{deluxetable}

\end{document}